\documentclass[%
 ,twocolumn%
 ,secnumarabic%
,amssymb, amsmath,nobibnotes, aps, pre]{revtex4}
\usepackage[T1]{fontenc}
\usepackage[cp1250]{inputenc}
\usepackage[polish,english]{babel}
\usepackage{graphicx}
\usepackage{bm}
\usepackage{color}

\begin{document}

\title{Effect of leader's strategy on opinion formation in networked societies}
\author{Paweł Sobkowicz}
\email{pawelsob@poczta.onet.pl}
\date{\today}

\begin{abstract}
The work investigates the influence of  leader's  strategy on opinion formation in artificial networked societies. The strength of the social influence is assumed to be dictated by distance from one agent to another, as well as individual strengths of the agents. The leader is assumed to have much greater resources, which allows him to tune the way he influences the other agents. We study various strategies of using these resources to optimize the conditions needed to `convince' the whole society to leader's opinion. The flexibility of the model allows it to be used in studies of political, social and marketing influence and opinion formation.
 \end{abstract}

\maketitle

\defcitealias{sobkowicz-lea1}{Paper~1}

\section{Modeling opinion formation in agent based computer models}

Our previous study (Paper I, \cite{sobkowicz-lea1}) has presented results of computer simulations within the model first proposed by \citet{nowak90-1}, developed further by  \citet{nowak96-1} and   \citet{kacperski99-1,kacperski00-1}, \citet{holyst01-1}. The model concerns the formation of public opinion through interactions between individual members of the society, taking into account differences in receptiveness, strength of influence and preexisting biases. The original work \cite{nowak90-1} has shown, using computer simulations, that interesting macroscopic behaviour can result from simple microscopical interacting agent model. The recent advances in understanding the nature of many social systems, including human societies and associations such as scientific collaborations, information exchange forums and techniques (e.g. WWW sites), friendship and acquaintance networks or nets of  sexual contacts show remarkable network properties \cite{strogatz01-1,albert02-1, dorogovtsev02-1, dorogovtsev02-2, dorogovtsev02-3, newman00-1, newman03-1, newman03-3}. It is then natural to apply the network paradigm to the study of opinion establishment in societies. 

The basis for the models used here and in Paper I (following  \cite{nowak90-1,nowak96-1,kacperski99-1,kacperski00-1,holyst01-1}) is:
\begin{itemize}
	\item A set of $N$ interacting agents forms a society. Interactions take form of one to one activities.
	\item Each agent has, at a given time, his `opinion'. The global characteristics of the behaviour of this `opinion' within the society is the topic of the research. For simplicity we treat the opinion as binary $+1 / -1$ variable.
	\item Each agent is characterized by the \textbf{strength} of his possible influence on other agents. This allows to model situations of uneven distribution of influence.
	\item One of the agents (the leader) is assumed to have the strength of influence much greater than the rest of the agents.
	\item  Interaction between particular agents is governed by the strength of agents as well as the \textbf{social distance} between the interacting agents. The cumulative effect of the strength and social distance is called \textbf{social impact}.
	\item The agents interact and influence each other in turns, changing their opinion after each full turn of interactions take place. Agents are allowed to interact with themselves, this mimics the phenomenon of self-support, or inclination to hold agent's present opinion.
	\item The model allows for extra-social influence or bias, applied uniformly to all agents.
	\item The model may allow for the noise in communication and changing individual opinion by adding an equivalent of temperature to the simulations.
\end{itemize}

Paper I studied mainly the effects related to the network topology and social separation between the agents. In the current paper we propose a new formulation of the problem of opinion formation, allowing to model the finiteness of leader's resources and allowing different strategies. As a result, we aim at study of how the leader {\textbf should} act to maximize his influence in networked societies.

We use here almost exactly the same basic framework as in Paper I. The only change in the formulation is in the way the leader's influence is introduced.

Traditional model of Nowak et al. and its extension by Kacperski and Hołyst  has set leader's  $L$ strength $s_L$ at constant value. The direct influence of the leader on any other agent ($i$) was then given by 
\begin{equation}
I_{iL} = s_L m_{iL} \sigma_L,
\end{equation}
and the overall impact, including the external conditions ($h$) and interactions with other, non-leader agents is
\begin{equation}
	I_i = s_L m_{iL} + \sum_{j \ne L} s_j m_{ij} \sigma_j + h,
\end{equation}
where $\sigma_L$ was assumed equal to 1. This way of measuring the strength may be referred to as general leader charisma --- it does not use any resources of the leader, the overall effect has no limitations with increasing society size etc. For this reason the traditional model's correspondence to real-life situations is rather limited. The only `damping' of leader's influence was contained in the immediacies $m_{iL}$, which are constant and thus did not allow to introduce any `strategies' for the leaders.

We introduce here a generalized model, in which the leader is assumed to have finite amount of resources $W$ (which may be related, for example, to wealth or to time available to the leader to spend with the other members of the society). The leader may then use different strategies of distributing the resources to achieve different {\textbf effective} strengths in interactions with different members of the society. For example, the leader may `spend' his resources on influencing a small core of his immediate neighbours, neglecting the others. This would allow to keep a cluster of agents supporting leader's opinion even in very disadvantageous conditions. On the other hand the resources may be used to `convince' the key agents (e.g. the most connected agents in Albert-Barabási society), and use their influence to promote leader's opinion. The model is constructed in such a way, that `equal' distribution of resources by the leader among all agents reproduces the traditional model with equal strength $s_L$.

In our simulations we compare results of a few general strategies, with the aim of modeling the most effective use of resources to achieve specific aims. For example, one of the questions is  which of the strategies is better at securing predefined percentage of supporters within a society; another problem might be to establish which kind of strategy minimizes the amount of resources needed to convince the whole society to leader's opinion in the presence of unfavorable external conditions.

\section{Details of the model}
\label{sec:DetailsOfTheModel}

As mentioned above, our model followed almost exactly that of  Kacperski  and Hołyst. We use the notation conventions introduced in Paper I.

The leader resources $W$ may be divided into individual portions --- time slots for individual meetings of a politician, amount of money spent on particular form of advertising etc. To social impact of the leader on agent $i$ is generalized to
\begin{equation}
I_{iL} = s_{iL} m_{iL} \sigma_L,
\end{equation}
where $s_{iL}$ are \textbf{individual (directed) strengths} of leader's influence on agent $i$. 
To `achieve' the desired effect in interaction with the target agent  $i$  by increasing the strength $s_{iL}$, the leader has to `spend' appropriate amount of resources, $w_i$. Of course finiteness of resources imposes
\begin{equation}
\sum_{i \ne L} w_i \le W.
\end{equation}
As in Paper I it is useful to introduce here the rescaled values. The scaling is given by
\begin{eqnarray}
	s_{iL}^R & = & \frac{s_{iL}}{N \bar{s}}, \\
	h^R & = & \frac{h}{N \bar{m}\bar{s}} ,
\end{eqnarray}
with averages $\bar{m}, \bar{s}$ excluding the leader.

There are many possibilities of mapping   $s_{iL}^R$ to $w_i^R$. In our approach we look for a cost function ${\cal W}$:
\begin{equation}
w_i^R = {\cal W}(s_{iL}^R)
\end{equation}
 which would take the following conditions into account:
\begin{itemize}
	\item The model should reduce to the traditional model when the total rescaled wealth $W^R$ is divided into $N$ parts (strictly speaking $N-1$, but we assume $N \gg 1$). Thus ${\cal W}(s_L^R) = W^R/N$, where $s_L^R$ is the rescaled uniform leader's strength  of the traditional model.
	\item It should be increasingly expensive to achieve larger and larger values of the strength $s_{iL}^R$, the cost function ${\cal W}$ should be linear or supralinear. In our simulations we have tested two forms of the cost functions: linear and quadratic. Most of the results presented here are for linear ${\cal W}$.
	\item In the limit of $s_{iL}^R \to 0$ also the corresponding cost $w_i^R$ should vanish.
\end{itemize}
The two forms of the cost functions used in our simulations were:
\begin{equation}
	{\cal W}_{\text{lin}}(s_{il}^R) = s_{iL}^R,
\end{equation}
\begin{equation}
	{\cal W}_{\text{quad}}(s_{il}^R) = \frac{(s_{iL}^R)^2}{s_L^R}
\end{equation}

The total social impact on agent $i$, by all agents (including the leader and the agent itself), in terms of the rescaled values, is given by
\begin{equation}
	I_i = s_{iL}^R \frac{m_{iL}}{\bar{m}} B + \sum_{j \ne L} s_{j}^R \frac{m_{ij}}{\bar{m}} \sigma_j B + h^R B,
\end{equation}
where we have introduced a very useful quantity $B = N \bar{s} \bar{m}$, which is the maximum value of the background influence of all non-leader agents if they all have $\sigma_j = 1$. We have also assumed that the `normal' agents do not have individual resources and that their strength of interaction is independent of the traget, that is $s_{ij} \equiv s_j$. 

\subsection{Leader's strategies}
The flexibility of assigning $w_i^R$ and therefore $s_{iL}^R$ to different agents $i$ allows us to simulate different leader strategies. In our simulations we have proposed the following procedure
\begin{itemize}
	\item The amount of wealth that he has at his disposal is given by a reference traditional system of equal distribution of strengths and costs ($s_{iL}^R \equiv s_L^R$), $W=N s_L^R$. 
	\item The leader choses the agents he is going to concentrate his efforts upon. The agents are ordered according to their importance in the chosen strategy.
	\item To avoid `spending' his efforts without guaranteed results, the leader should adjust the individual strengths $s_{iL}^R$ in accordance with the immediacy value $m_{iL}$ and other conditions (e.g. social temperature $T^R$ and external influence $h^R$. This is achieved through  a threshold parameter $t$ which reflects the value of the leader strength toward agent $i$ needed (on the average) to convince this agent. One example of expression for $t$ is the threshold for strength needed to overcome the combined influence of external conditions ($h^R$) and maximum negative impact of all non-leader agents on $i$. In this case $t=1-h^R$. The leader then calculates the strength needed to pass this threshold $s^*_{iL} \geq t \bar{m} / m_{iL} $. In some cases smaller values of $t$ may be used, still ensuring desired effect of leader's interaction for targeted agents. For example, in random initial conditions it should be sufficient to use $t\geq  -h^R$.
	\item The leader interacts with the targeted agents in order of importance, each time using up the necessary amount of resources $w^{R*}_i = \cal{W}(s^*_{iL}$, until he runs out of resources. For all other agents the influence of the leader is assumed to be zero.
\end{itemize}

This procedure allows, for example, that at small values of the reference traditional uniform leader strength  ($s_L^R \ll -h^R$), for which we have shown in Paper I that the support for the leader is negligible, to concentrate on a few agents to obtain a `cadre' of followers. Common sense suggests that spending the same amount of resources on 1/10th of targets allows, on average, 10 times greater per target expense, and following this increase in expected individual results. While the increase may not be linear, it is worth remembering that the traditional model used resources proportional to the number of agents in a society.

The key for leader's strategy lies in the way the target agents are chosen and ordered. We propose here a few  natural candidates for such strategies

\begin{description}
	\item[Neighbours first.] As we follow Paper I in the way the immediacies $m_{ij}$ are calculated from network distances between agents $i$ and $j$, the values of $m_{iL}$ decrease with decreasing distance from the leader. It follows that the cost of convincing one's close neighbors is the least. The total wealth can be spend most effectively on the closest neighbours, minimizing the expenses on `lost cases' --- agents so remote that the expense needed to fulfill condition for $s^*_{iL}$ is prohibitive. In this way, the leader can assure surpassing the support threshold for a limited cluster of his neighbours. In terms of the spatial model of Nowak, this corresponds to influencing a limited circle around the leader. It is natural to expect that due to concentration of effort, the size of the bubble (or support cluster size in abstract space networks) would be greater than in the reference traditional case.
	\item[Convincing the highly connected agents first.] In Paper I we have shown that in scale free networks of Albert-Barab\'asi, if the leader is in the highly connected node his influence on other agents is much more pronounced. Our directed effort model allows the leaders who are not in highly connected nodes to proceed as follows: spend as much as necessary on the highly connected agents (regardless if they are in close neighbourhood or not), and then count that the combined influence of these agents would serve as a vehicle of leader's opinion.
	\item[Mixed strategy: top influencers plus the rest.] A combination of the  previous strategies. The targets of leader's attention are just a few of the most important agents (defined as in previously presented strategies), and then the remaining resources are spent on all other agents. This allows the leader to be sure that while no agent would be entirely without leaders influence (e.g. through media), the key actors would be `personally' contacted with appropriate resources. We have investigated three variants of this strategy, with emphasis on the closest neighbours, most connected or strongest agents.
\end{description}

\subsection{Results}
\label{sec:Results1}

\definecolor{darkgreen}{rgb}{0,.35,0}

The first striking contrast between the results of any directed strategy simulation and the results of the traditional approach, is linear growth of the percentage of the population that the leader's supporters form with the increase of the resources available, measured by $s_L^R$. This is due to linear growth in the number of agents `personally influenced and convinced'. While in the traditional model, for $s_L^R \ll -h^R$ the number of supporters was near zero, here, due to the way the resources are used, the leader can `guarantee' the support of the agents he uses the resources on. In our simulations we have uset the threshold $t = 1.2 (1-h^R)$, which according to simulations presented in Paper~I is sufficient to convince the targeted agent.  The simulations recreate thus observations from real life, where dedicated, close relationships of average people  are sufficient to establish  small but loyal groups of supporters.
As long as the threshold $t$ is large enough to ensure the conversion of the targeted agent, the support fraction $f$ is simply given by the number of agents the leader has resources for. For large range of values of $s_L^R$ the growth of $f$ is linear, reaching 1 when the resources allow to contact everyone in the population.

For mixed strategies, we did not observe any new or interesting phenomena. Apart from the small group of directly influenced agents, the support fraction behaved exactly in the same way as in traditional model. 

Summarizing the effects of directed resources strategies, one can state that they prove themselves in influencing and converting this part of the population that they are aimed at, but hardly matter for the rest of the society.

\section{Strategies with resource transfer}
\label{sec:restransfer}

As we have shown, for large populations, the ability to ensure the conviction of a part of the population,  does not bring out results going beyond the directly approached part of population. The reason is quite simple: due to enormous asymmetry between the social impact of the leader and \textbf{individual} impacts of any of the other players (with the ratio given by $N \bar{s}$), even targeting the most connected or the strongest agents does not produce enough momentum for the leader's cause. The influence of a single non-leader agent is simply too small. 

Let's consider now a new approach, allowing for entirely new kind of strategy. In the new model, the leader can not only \textbf{direct} his resources at will to influence chosen groups of agents, but also can \textbf{transfer some of the resources} at his disposal to selected agents. This would correspond to, say, establishing local headquarters of leader's party, with local media funds or to investment in training of leader's representatives, who would then `substitute' for the leader himself. The question that arises is: can the leader by such procedure of distributing his resources increase the rate of conviction and achieve the state of global supportiveness faster or cheaper? 

The strategy of the leader is described through a single parameter $N_{CL}$ --- number of agents that the leader wants to turn into \textbf{co-leaders} of his cause. The amount of the resources needed to ensure the support of the co-leaders is deducted from the total resources $W$, and the remainder is divided, for example equally, among the group of newly formed co-leaders (including the leader itself). The social influence on other agents is then calculated in the way similar to the traditional strategy described in previous section, but for every co-leader separately. 

One can expect, that due do the choice of the agents that form the co-leader team (for example the most highly connected agents, whose proximity to any other agent is comparatively small) the combined effect of the same amount of resources $W^R$, applied through multiple actors would be much more effective than the case of the single leader. This is indeed observed: for a given $h^R$, the threshold of the transition to supportive state, $s_L^*$ is shifted to smaller values.

Figure~2 compares the fraction $f$ of agents supporting the leader as function of $s_L^R$ (which is a convenient way of presenting the resources available to the leader, $s_L^R = W^R / N$, directly comparable to traditional model). The results were calculated for a given value of unfavorable external conditions ($h^R=-1.5$), and for random initial opinion distribution ($\sigma=0$). The network used has the Albert-Barab\'asi topology, with the leader in random position. The three sets of results correspond to three values of the number of co-leaders $N_{CL}$. We compare the results with the simulations in traditional model, for the leader in a random position, and for the leader in the most highly connected positon.

As noted in Paper~I, due to enormous difference between the number of connections for typical agent and for the highly connected (HC) agents, there is significant difference in the threshold value of the leader strength $s^*_L$ at which the population reaches supportive state. The advantage of starting from highly connected position is obvious. Our new resource transfer model allows the typical agent --- once it has appropriate resources --- to obtain results similar to those `reserved' for the HC leaders in traditional model.

Figure~3 presents the values of the threshold strength $s^*_L$ as function of the number of co-leaders chosen by the original, randomly placed leader to help him. We present results for bot random (unbiased $\sigma=0$) and negatively biased ($\sigma=-1$) starting conditions. In both cases one can determine a range of values of $N_{CL}$ for which the  cooperation  results in far better results tahn in the traditional model. The horizontal lines correspond to threshold values for the traditional model for the same random leader position and for the most highly connected leader. For the case of random initial opinion, proper choice of co-leaders, their conversion and later distribution of resources results in effective change of the threshold almost to the value obtained for the best connected leader position.

It is interesting to note, that if initially  leader is in the highest connected position, application of our strategy does not bring visible improvement. Any co-leader would have less optimal position than the original one, thus multiplication of the number of channels to an average agent does help the leader's cause.

Another interesting application of the resource-transfer strategy is the nearest neighbour (NN) network. In Paper~I we have shown, that due to the rapid growth of separations of agents in NN network and its highly localized nature, the support for the leader grows slowly and linearly with increasing $s_L^R$. The resource transfer model allows the leader to use a new strategy aimed at effective shortening of these distances. The leader can convert agents dispersed evenly in the society and later redistribute its resources to these agents. As the distance from an  average agent to the closest co-leader is now much smaller, the social impact is greater and instead of the linear growth we observe a faster ramp-up of $f$ as function of $s_L^R$ and transition to fully supportive state. The shape of this ramp-up depends on the number of co-leaders $N_{CL}$, examples are presented in Figure~4. As in the case of the AB networks, the application of the new strategy leads  to easier achievement of the fully supportive state. Figure~5 presents the values of the threshold $s^*_L$ as function of $N_{CL}$ for random and negative initial conditions. Because the `nature of the task given to co-leaders' is here different from the AB network case (namely, to shorten the distances between leaders and average society members)  the optimum values of  $N_{CL}$ are different, but the pattern remains similar: application of new strategy has clear advantage for the leader.

There are other situations where application of resource transfer strategy would result in immediate improvement of the ability to achieve a supportive state. Good example is provided by all networks in which there are communication bottlenecks: regions of networks connected by few links and agents with high `betweenness'. In such situation, establishment of `local representatives' with appropriate resources is crucial condition for success.

The idea of directed, concentrated application of available resources and their re-distribution is drawn from real life examples: from politics to sales activities. Examples of `think globally, act locally' are too numerous to present here. We think that our model, simple as it is, offers a framework for the analysis of real-life opinion formation extending significantly beyond the traditional approach.


\begin{figure*}[!h]
\centering
\includegraphics[height=16cm]{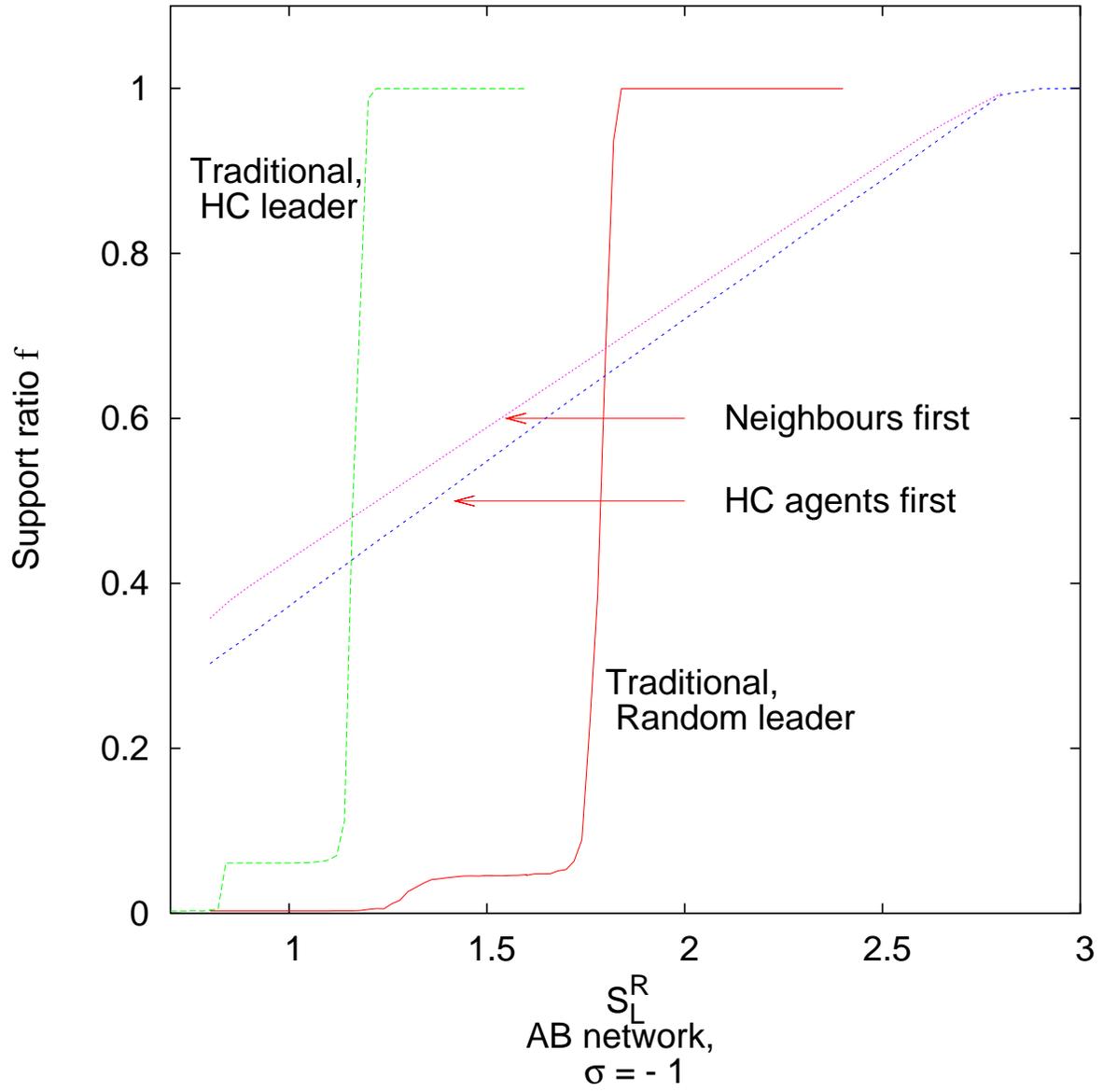}
\caption{Support fraction $f$ as function of leader strength $s_L^R$ for the `neighbours first' and `highest connected agents first' strategies, compared to traditional models. Albert-Barab\'asi network, $h^R=-1.5$, $t=1.2*(1-h^R)$.
\label{figure1}}
\end{figure*}

\begin{figure*}[!h]
\centering
\includegraphics[height=16cm]{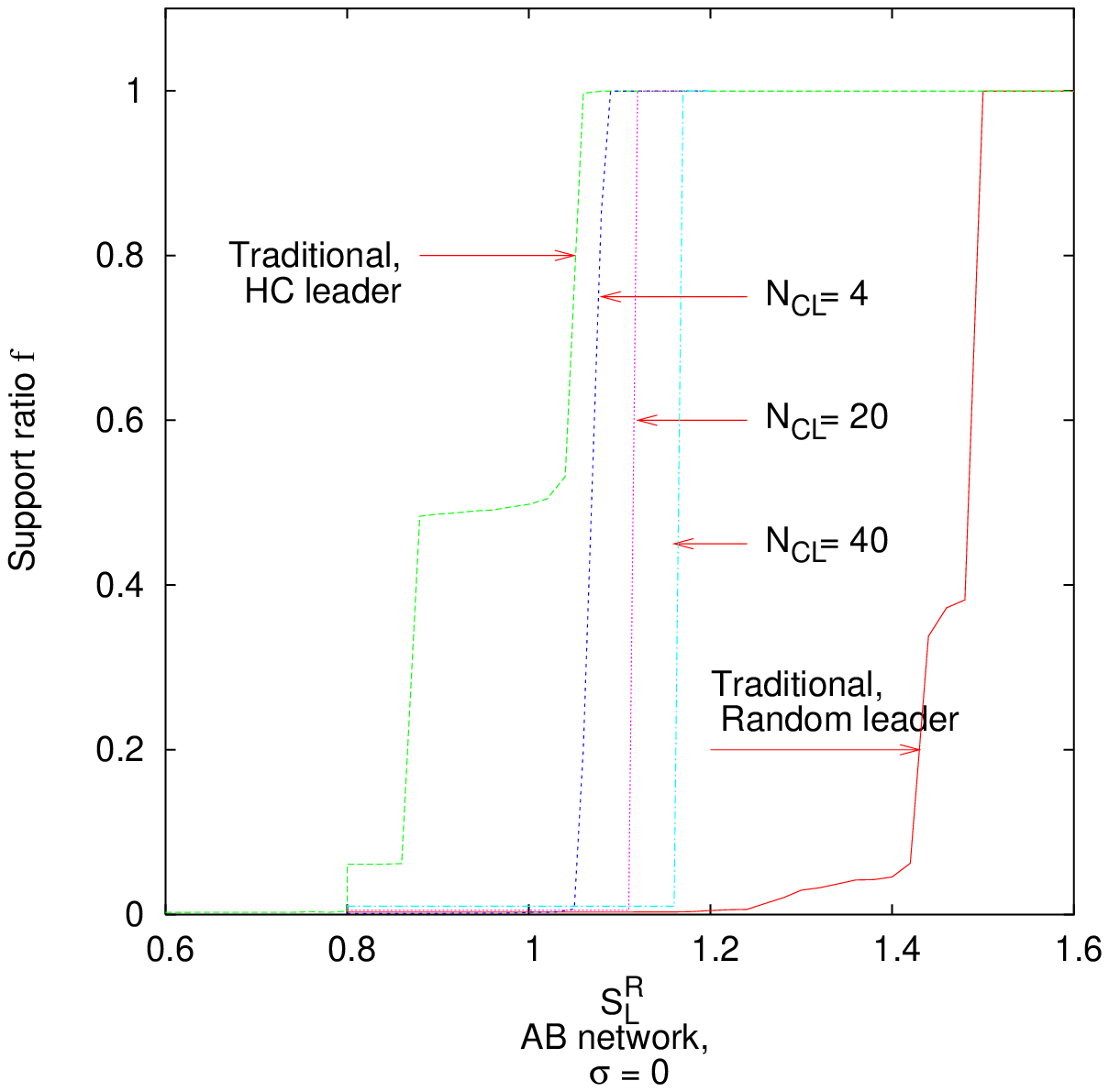}
\caption{Support fraction $f$ as function of leader strength $s_L^R$ for the  strategy with resource transfer for three values of the co-leader number $N_{CL}$, compared to traditional models. Albert-Barab\'asi network, $h^R=-1.5$, $t=1.2*(1-h^R)$, average $\sigma = 0$, leader in randomly chosen position.
\label{figure1}}
\end{figure*}

\begin{figure*}[!h]
\centering
\includegraphics[height=12cm]{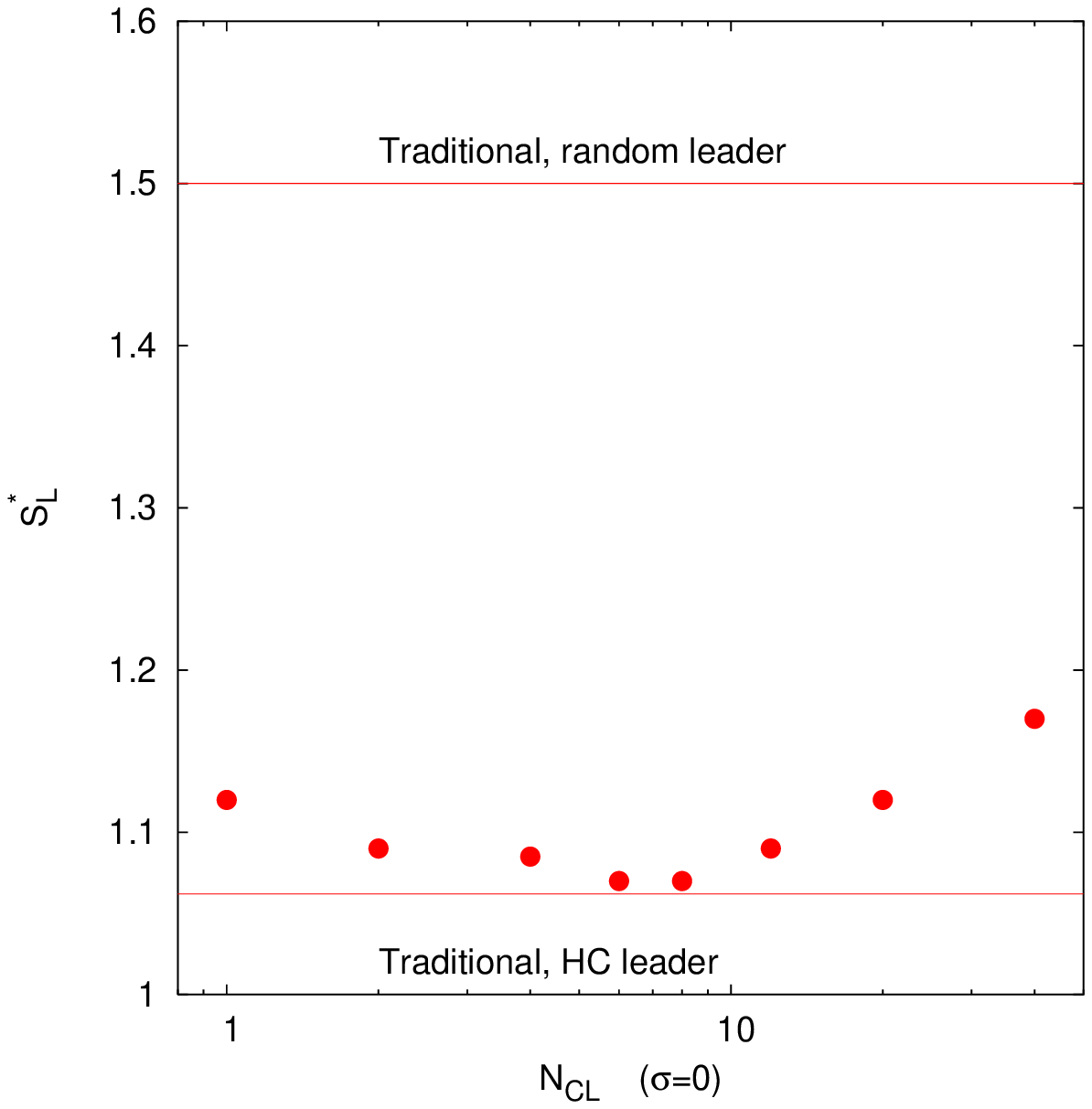}
\includegraphics[height=12cm]{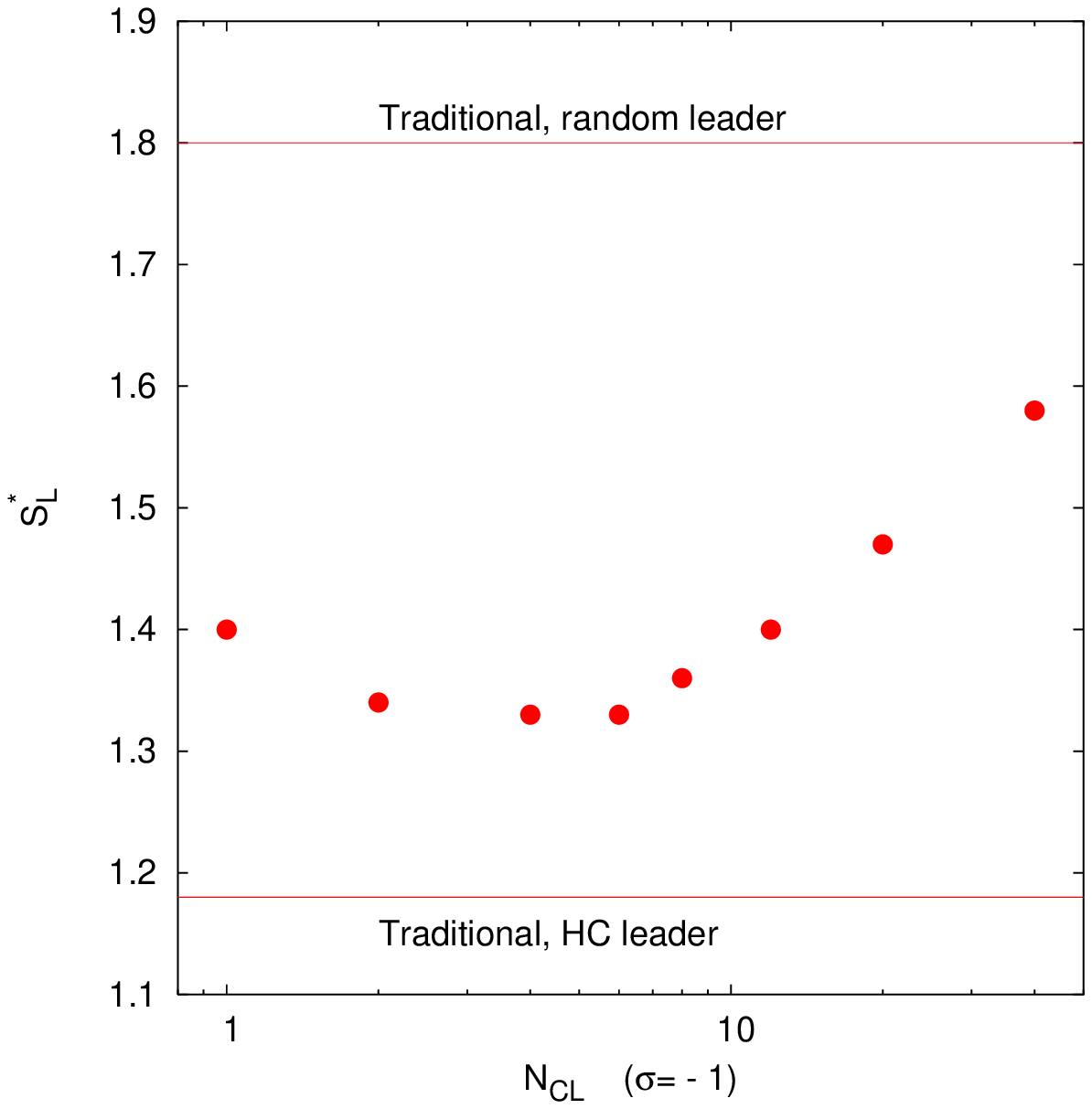}
\caption{The rescaled leader threshold strength value $s^*_L$ at which the society reaches fully supportive state, as function of the number of co-leaders in strategy with resource transfer. Albert-Barab\'asi network, $h^R=-1.5$, $t=1.2*(1-h^R)$. Upper panel: random initial distribution of $\sigma_i$, lower panel disadvantageous starting condition $\sigma_i\equiv -1$. Horizontal lines show the threshold values for traditional simulations (with constant leader strengh) for leader in random position and in the most highly connected position.
\label{figure1}}
\end{figure*} 

\begin{figure*}[!h]
\centering
\includegraphics[height=16cm]{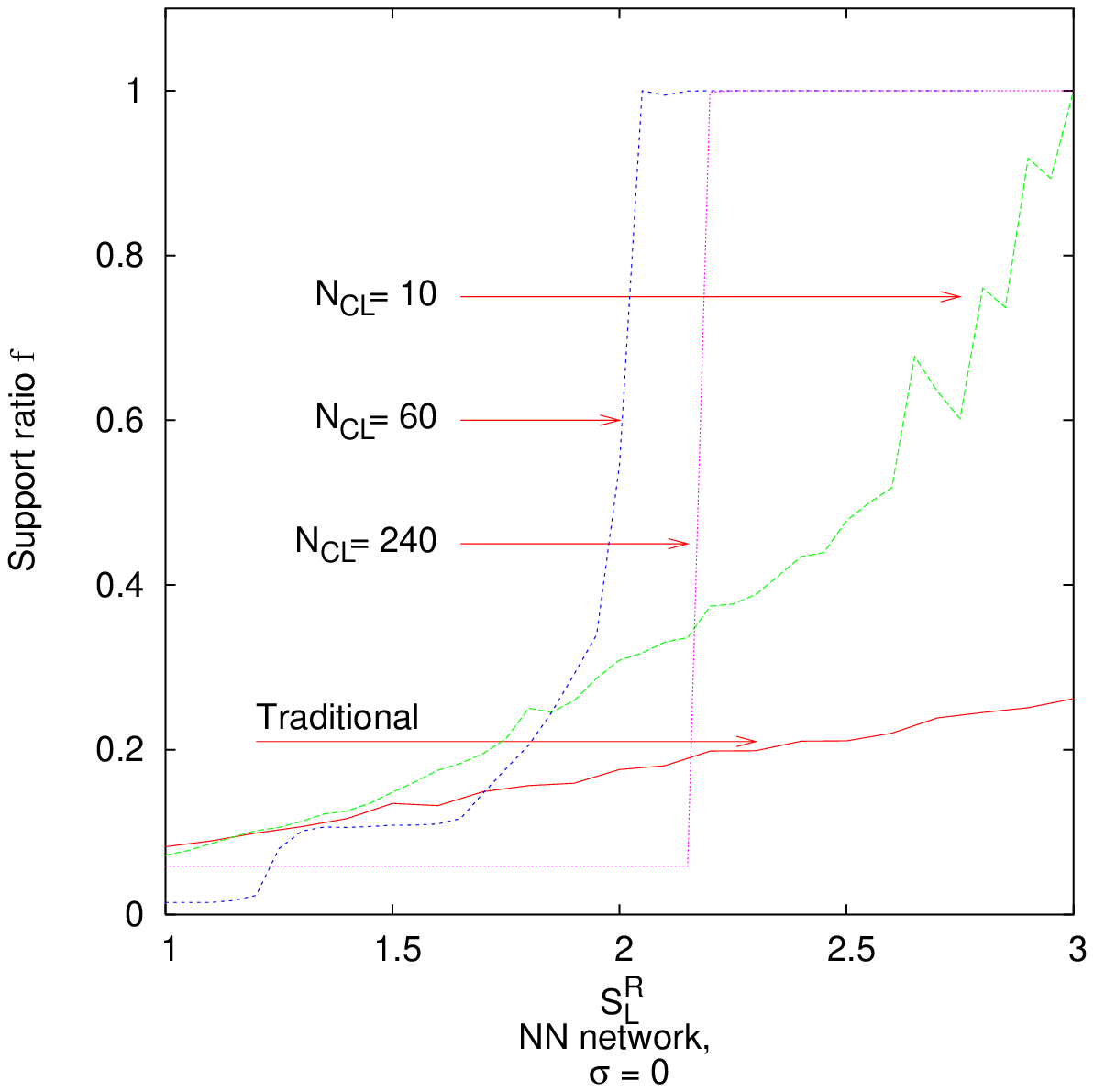}
\caption{Support fraction $f$ as function of leader strength $s_L^R$ for the  strategy with resource transfer for three values of the co-leader number $N_{CL}$, compared to traditional models. Nearest Neighbour network, $h^R=-1.5$, $t=1.2*(1-h^R)$, average $\sigma = 0$, leader in randomly chosen position.
\label{figure1}}
\end{figure*}

\begin{figure*}[!h]
\centering
\includegraphics[height=12cm]{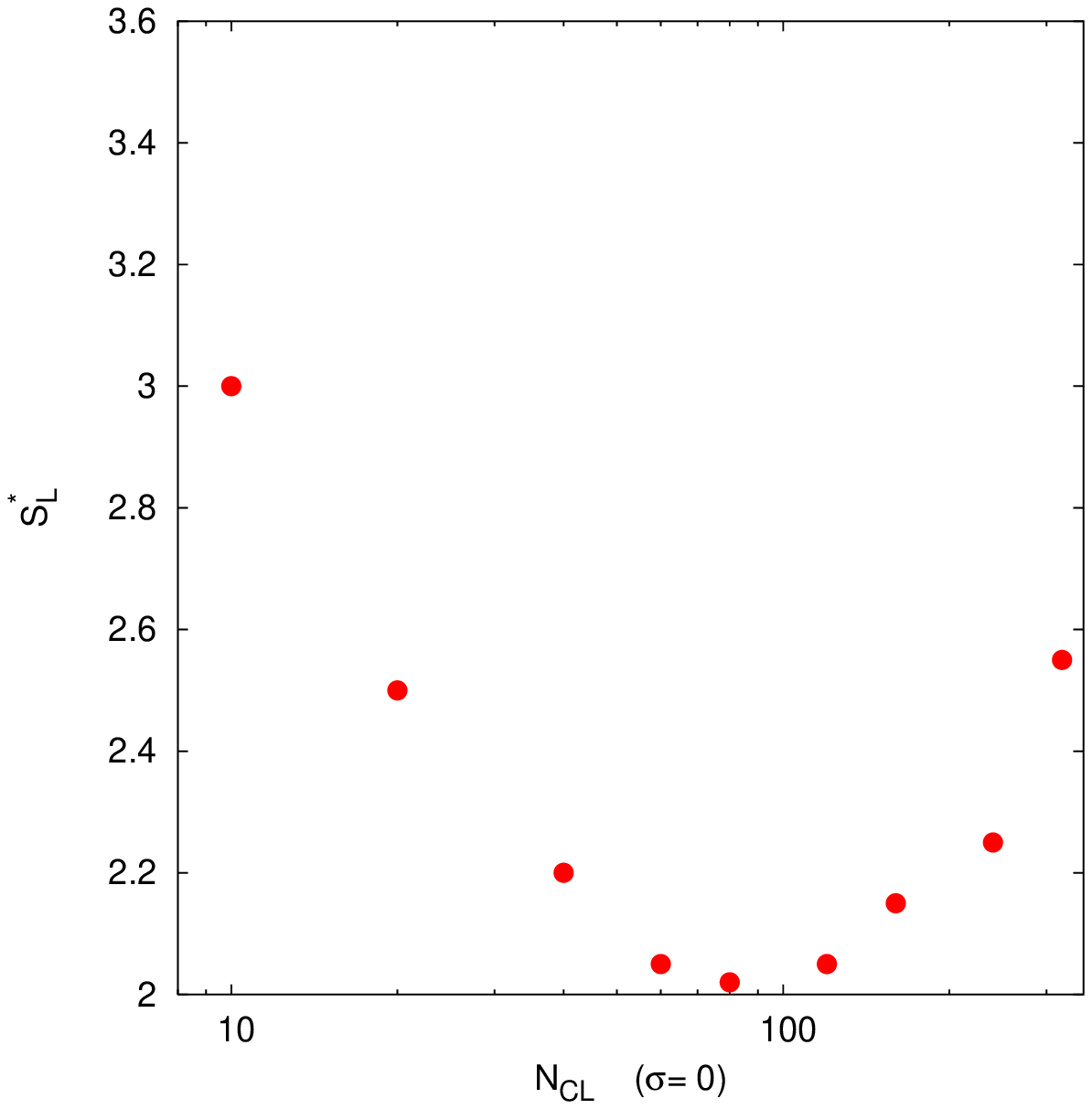}
\includegraphics[height=12cm]{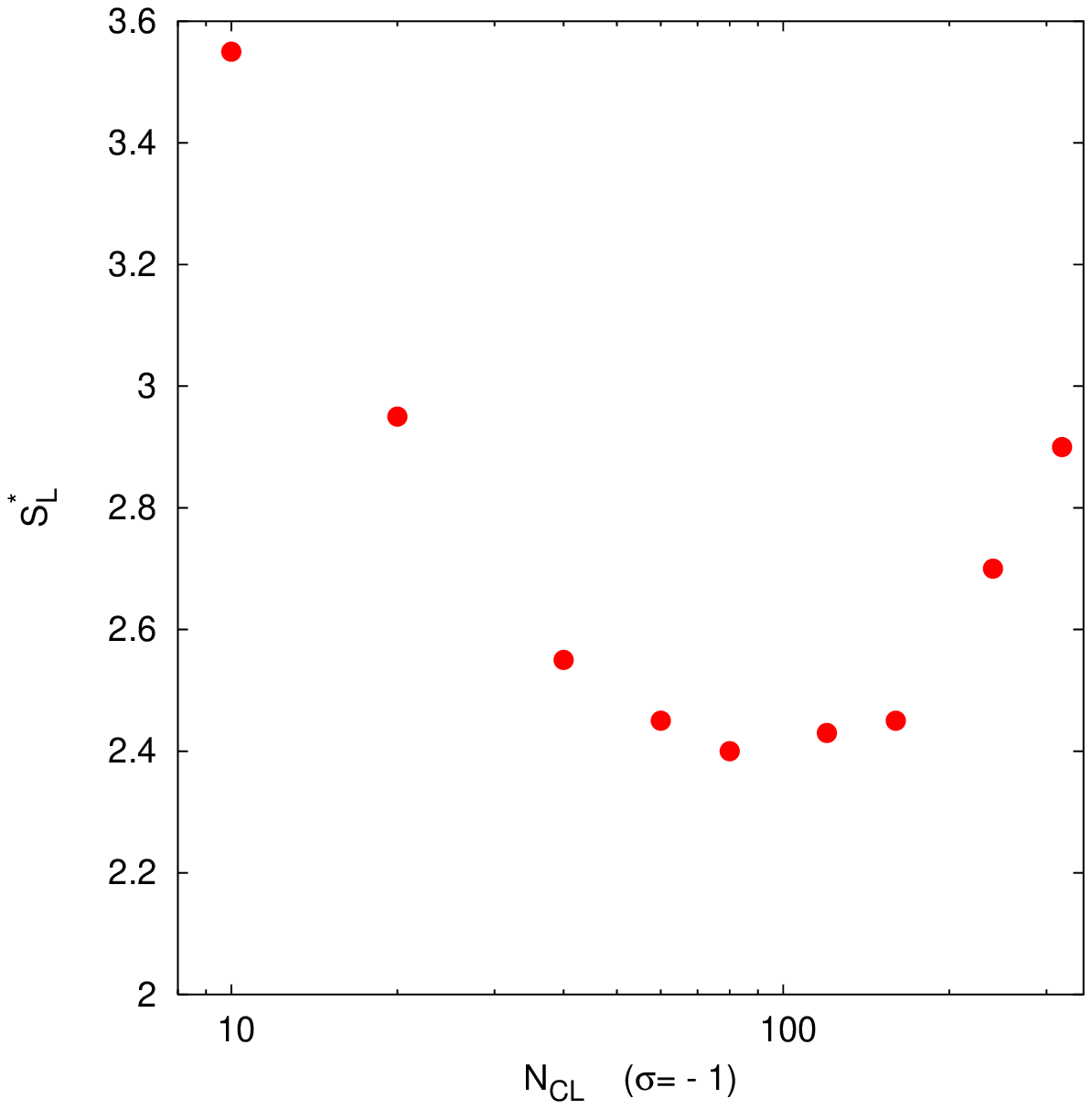}
\caption{The rescaled leader strength value $s^*_L$ at which the society reaches fully supportive state, as function of the number of co-leaders in strategy with resource transfer. Nearest Neighbour network, $h^R=-1.5$, $t=1.2*(1-h^R)$. Upper panel: random initial distribution of $\sigma_i$, lower panel disadvantageous starting condition $\sigma_i\equiv -1$.
\label{figure1}}
\end{figure*}

 \end{document}